# The Efficiency of Ideal Thermomagnetic Engine

A.G. Kvirikadze, M.D. Zviadadze, V.A. Barnov, G.G. Kiknadze, L.A. Zamtaradze

I. Javakhishvili Tbilisi State University, 3 Chavchavadze Ave. Tbilisi, 0128, Georgia

E. Andronikashvili Institute of Physics, 6 Tamarashvili Str., Tbilisi, 0177, Georgia

zviadadzemichael@yahoo.com

## Abstract

The article contains calculation of the efficiency of the ideal thermomagnetic engine whose working body is ferromagnetic has been suggested. Suggested method being applicable to any heat engine operating on the basis of structural phase transitions in solid bodies.

# 1. Introduction

Recently, in connection with the ever-increasing world-wide energetic crisis a great attention has been paid to heat engines operating at low-potential energy. Among them are the engines operating at using the structural phase transitions in solid bodies: martensite-austenite, ferroelectric-dielectric, ferromagnetic-paramagnetic.

The presented work gives the calculation of the efficiency of the ideal thermomagnetic engine (TME), the working body of which is a ferromagnetic material.

# 2. Basic physics of a work of the thermomagnetic engine

As it is well known, the magnetic moment $\vec{M}$ located in nonhomogeneous magnetic film $\vec{H}$ is acted not only by the precessional force proportional to the vector product $[\vec{M} \times \vec{H}]$, but also by the force $\vec{F} = (\vec{M}\vec{\nabla})\vec{H}$ caused by the inhomogeneity of magnetic field ($\vec{\nabla}$ is the gradient operator). For the ferromagnetic material of the final dimensions, $\vec{F}$ force is equal to the sum of forces acting on different sections of the sample. It is assumed that this force is proportional to the full magnetic moment. Besides, it is assumed that the gradients of magnetic field are such that we can neglect the magnetostrictive effects and, hence, not take into account the changes of the shape and the volume of sample. In TME, just this force $\vec{F}$ does the work and therefore, to achieve the high values of this force, the magnetic fields with a high gradient and the ferromagnetic materials with a large magnetic moment should be chosen.

The domain structure of ferromagnetic material is significantly determined by the demagnetizing field that depends on the shape of sample [1]. Two cases are known, when demagnetizing field is equal to zero and the sample is in the single-domain state. The sample made of uniaxial ferromagnetic material is a parallel-sided infinite plate with planes parallel to the magnetization vector, and an infinite cylinder with axis parallel to the magnetization vector.

Besides, as the limited dimensions of the region, where the external magnetic field is created, in order the resultant force at the input and at the output be maximum, it is necessary to change significantly the magnetic properties of sample within the range of magnetic field. These properties are changed most strongly at the phase transitions of the second order – ferromagnetic-paramagnetic – in the vicinity of Curie temperature $T_C$ [2].

Though the dependence $M(T)$ in the vicinity of $T_C$ has a rather complicated form, it is important that a sharp change of magnetic moment takes place in the narrow temperature band $\Delta T$.

Thus, if the temperature of sample at the input is $T_1 < T_c - \Delta T$, the magnetic moment $M(T_1)$ is of $M_0$ order (the maximum magnetic moment of sample). If we heat the sample up to the temperature $T_2 > T_C + \Delta T$ when it passes the magnetic field then, its magnetic moment is $M(T_2) = M_p(T_2, H_{ef})$ when it leaves the magnetic field, where $M_p$ is the magnetic moment of paramagnetic material, $H_{ef}$ is the certain efficient magnetic field at the output in the region of strong inhomogeneity of the field. Obviously, $M_p(T_2,0) = 0$. In the general case, $M(T_2)$ depends in a complicated way on the temperature $T_2$ and on the character of motion that determines $H_{ef}$. The resultant force depends on the difference $M(T_1) - M_p(T_1, H_{ef})$, and just this difference determines the character of motion of sample, i.e. the system with feedback is obtained, the properties of which depend significantly on the method of heating the sample. As it was already mentioned, due to the small characteristic dimensions of the regions of magnetic field, in the case of moving sample a detailed study of the processes of heat transfer from the heater to the sample is necessary. The processes of heat transfer are less limited in connection with the less strict requirements for the cooling regions. The processes of heat transfer become more important taking into account the cyclic operation of the engine. In this case, the only way to act on the system from the outside is a thermal contact. As it is known [2], the process of heat transfer is reversible at neglecting the magnetostrictive effects; therefore, to obtain the significant difference in the quantity of heat generated by the heater and transferred to the cooler, it is necessary to change the conditions of heat transfer. In this case, the only method of changing these conditions is to heat and to cool the sample in different magnetic fields.

## 3. Calculation of the efficiency of ideal thermomagnetic engine

Let us denote the quantity of heat necessary for heating the sample from $T_1$ to $T_2$ in homogeneous magnetic field $H$ by $Q_H$, and by $Q_0$ the quantity of heat transferred by this sample to the cooler at cooling from $T_2$ to $T_1$ in the magnetic field equal to zero. Then, the difference $Q_H - Q_0$ is equal to the work done by the magnetic field, and the value

$$\eta = \frac{Q_H - Q_0}{Q_H} \tag{1}$$

is the efficiency.

Thus, the efficiency depends significantly on the engine design, of the configuration of external magnetic field, on the conditions i.e., where the heating is started and at which fields, it is desirable to carry out the heating in such sections of the magnetic field where it has the maximum value and is homogeneous, etc.

To estimate the efficiency of TME in most interesting temperature region $T_1 \ll T_C \ll T_2$, we can make use of the idealized pattern, assuming that the ferromagnetic sample is heated uniformly in homogeneous magnetic field and, thus, is cooled in the absence of magnetic field. Taking into account the reversibility of thermal contact, and neglecting the magnetostrictive and magnetocaloric effects, as well as the thermal expansion, $Q_H$ and $Q_0$ can be easily expressed by the change of internal energy of the sample. Indeed, under the proposed conditions

$$Q_H = E(T_2, H) - E(T_1, H), \tag{2}$$

where $E(T_2, H)$ and $E(T_1, H)$ are the internal energies of ferromagnetic material in the magnetic field $H$ at the corresponding temperatures. Hence, the heat transferred from the sample to the cooler is

$$Q_0 = E(T_2, 0) - E(T_1, 0). \tag{3}$$

Internal energy of ferromagnetic crystal can be presented in the form of the sum

$$E(T, H) = E_m(T, H) + E_l(T), \tag{4}$$

where $E_m(T,H)$ is the internal energy of magnetic subsystem of ferromagnetic material, and $E_l(T)$ is the internal energy of the crystal lattice not depending on the magnetic field owing to the fact that the magnetostrictive effects are neglected.

To estimate the internal energy of the magnetic subsystem, we can use the simplest form of Hamiltonian

$$H = -\frac{1}{2}\sum_{l \neq n} J_{ln} \vec{S}_l \vec{S}_n - 2\mu_0 H \sum_l S_{zl}, \tag{5}$$

where the exchange integral $J_{ln}$ is considered to be different from zero only for the nearest neighbors, $\mu_0 = \frac{e\hbar}{2mc}$ is the Bohr magneton, and $\vec{S}_l$ is the spin operator for the $l$-th atom.

Below the narrow region of phase transition, the internal energy of ferromagnetic material can be presented as a sum of the energy of ground state $E_m(0,H)$ and of the energy of magnon gas $U_m(T,H)$.

$$E_m(T,H) = E_m(0,H) + U_m(T,H). \tag{6}$$

The energy of ground state $E_m(0,H)$ for magnetic subsystem with Hamiltonian (5) is

$$E_m(0,H) = V\left\{ n\frac{S^2}{2}\sum_{l \neq 0} J_l M_0 H \right\}, \tag{7}$$

where $M_0$, the density of magnetic moment of the ground state $M_0 = \frac{m}{V} = \frac{2\mu_0 NS}{V}$ is determined by the maximum value of the spin of separate atom and by the density of atoms $n = \frac{N}{V} \sim \frac{1}{v_0}$ ($v_0$ is the volume of elementary cell per one magnetic atom, and $v_0 \sim \frac{1}{a^3}$, where $a$ is the lattice constant).

In expression (7) the summation over $l$ is made only over the nearest neighbors of the magnetic atom located at the origin of coordinates, and $\sum_{l \neq 0} J_l \approx zJ$, where $z$ is the number of the nearest neighbors and $J \sim \xi \frac{e^2}{a}$ is the exchange integral, $a$ is the dimensionless parameter, $\xi$ for typical ferromagnetic materials is of the order of $\xi \sim 0.1 - 0.04$. According to the order of magnitude $J \sim T_C$.

Magnon energy with the wave vector $k$ for the system with Hamiltonian (5) is

$$E(k) = S\sum_l J_l (1 - \cos k\vec{r}_l) + 2\mu_0 H \tag{8}$$

and in the long-wave approximation, for the crystal with cubic symmetry we have:

$$U(k,H) = \alpha k^2 + 2\mu_0 H; \quad \alpha = \frac{Sa^2}{2}\sum_{l \neq 0} J_l. \tag{9}$$

Thus, the internal energy of magnon gas is equal to

$$U_m(T,H) = V \frac{1}{(2\pi)^3} \int \frac{E(k,H)}{\exp(E(k,H)/k_B T) - 1} d^3k. \qquad (10)$$

Then, for the internal energy of ferromagnetic material in paramagnetic state at $T_2 \gg T_C$ we have

$$E_m(T_2, H) = -\frac{1}{2} \sum_{l \neq n} J_{ln} \vec{S}_l(T_2, H) \vec{S}_n(T_2, H) - 2\mu_0 H \sum_l S_{zl}(T_2, H), \qquad (11)$$

besides, the relation between the density of magnetic moment $\vec{M}(T,H)$ and the average value of spin $\vec{S}_{zl}(T_2, H)$ is of the same character as in case of the ground state, i.e.

$$\vec{M}(T,H) = 2\mu_0 \frac{N}{V} \vec{S}(T,H). \qquad (12)$$

For this reason, at $T_2 \gg T_C$, because of the infinitesimal of $M(T_2, H)$, we can assume that

$$E_m(T_2, H) \sim E_m(T_2, 0) \sim 0$$

thus, the heat absorbed by the ferromagnetic material at heating from $T_1$ to $T_2$ in the magnetic field is equal to

$$Q_H = E_l(T_2) - E_l(T_1) + V \left\{ n \frac{S}{2} \sum_{n \neq 0} J_n \right\} - V \frac{1}{(2\pi)^3} \int \frac{U(k,H)}{\exp(U(k,H)/k_B T) - 1} d^3k. \qquad (13)$$

The heat transferred to the cooler at cooling the sample from $T_2$ to $T_1$ in the magnetic field equal to zero is absolutely the same

$$Q_0 = E_l(T_2) - E_l(T_1) + V \frac{nS^2}{2} \sum_{n \neq 0} J_n - V \frac{1}{(2\pi)^3} \int \frac{U(k,H)}{\exp(U(k,H)/k_B T) - 1} d^3k \qquad (14)$$

hence, the work performed by the system is

$$Q_0 = E_l(T_2) - E_l(T_1) + V \frac{nS^2}{2} \sum_{n \neq 0} J_n - V \frac{1}{(2\pi)^3} \int \frac{U(k,H)}{\exp(U(k,H)/k_B T) - 1} d^3k, \qquad (15)$$

and the efficiency of TME, according to (1), (13), (15) is

$$\eta = \frac{M_0 H + \frac{1}{(2\pi)^3} \int \left( \frac{U(k)}{\exp(U(k)/k_B T_1) - 1} - \frac{U(k,H)}{\exp(U(k,H)/k_B T_1) - 1} \right) d^3k}{M_0 H + \frac{nS^2}{2} \sum_{n \neq 0} J_n - \frac{1}{(2\pi)^3} \int \frac{U(k,H)}{\exp(U(k,H)/k_B T_1) - 1} d^3k + \frac{E_l(T_2) - E_l(T_1)}{V}}. \qquad (16)$$

The difference in internal energy of unit volume of crystal lattice can be expressed by the difference in energy of phonon gas of unit volume or by specific heat capacity of the lattice

$$\frac{E_l(T_2) - E_l(T_1)}{V} = \frac{1}{(2\pi)^3} \int U(k) \left( \frac{1}{\exp(U(k)/T_2) - 1} - \frac{1}{\exp(U(k)/T_1) - 1} \right) d^3k = \int_{T_1}^{T_2} C_l(T) dT . \quad (17)$$

Thus, the efficiency equals to

$$\eta = \frac{M(T_1)H}{M(T_1)H + \frac{nS^2}{2} \sum_{n \neq 0} J_n - \frac{1}{(2\pi)^3} \int \frac{U(k,0)}{\exp(U(k,0)/k_B T_1) - 1} d^3k + \int_{T_1}^{T_2} C_l(T) dT} . \quad (18)$$

It should be noted, that on the basis of the approximations made above, expression (18) is the efficiency of the ideal TME, therefore, the estimation (18) can be considered as a rough approximation for the efficiency of real TME.

**Acknowledgements:** This work is supported by the Shota Rustaveli National Science Foundation, Grant GNSF 712/07. The authors are grateful to M. Nikoladze for help with processing articles